\documentclass[AMA,LATO1COL]{WileyNJD-v2}
\usepackage[T1]{fontenc}
\usepackage[utf8]{inputenc}
\usepackage{tgbonum}
\usepackage{multicol, blindtext}
\usepackage{amsmath,amssymb,amsfonts}
\usepackage{svg}
\usepackage{textcomp}
\usepackage{xcolor}
\usepackage{tikz}
\usepackage[figuresright]{rotating}
\usepackage{diagbox}
\usepackage{multirow}
\usepackage{colortbl}
\usepackage{colorwav}
\usepackage{etoolbox}
\usepackage{graphicx}
\usepackage{caption} 
\newcommand\BibTeX{{\rmfamily B\kern-.05em \textsc{i\kern-.025em b}\kern-.08em
T\kern-.1667em\lower.7ex\hbox{E}\kern-.125emX}}

\articletype{Research Paper}%

\received{ }
\revised{ }
\accepted{ }

\def\BibTeX{{\rm B\kern-.05em{\sc i\kern-.025em b}\kern-.08em
    T\kern-.1667em\lower.7ex\hbox{E}\kern-.125emX}}

\usetikzlibrary{shapes}
\usetikzlibrary{positioning,arrows}
\usetikzlibrary{patterns}
\renewcommand*{\do}[1]{
  \storeRGBofWavelength{\Rval}{\Gval}{\Bval}{#1}
  \definecolor{wl#1}{rgb}{\Rval, \Gval, \Bval}
}
\docsvlist{380,390,400,410,420,430,440,450,460,470,480,490,500,510,520,530,540,550,560,570,580,590,600,610,620,630,640,650,660,670,680,690,700,750,780}

\pgfdeclarehorizontalshading{spectrum}{2cm}{
  color(0cm)=(wl380);
  color(0.1cm)=(wl390);
  color({0.2cm})=(wl400);
  color({0.3cm})=(wl410);
  color({0.4cm})=(wl420);
  color({0.5cm})=(wl430);
  color({0.6cm})=(wl460);
  color(0.7cm)=(wl450);
  color(0.8cm)=(wl460);
  color(0.9cm)=(wl470);
  color(1cm)=(wl480);
  color(1.1cm)=(wl490);
  color(1.2cm)=(wl500); 
  color(1.3cm)=(wl510);
  color(1.4cm)=(wl520);  
  color(1.5cm)=(wl530);
  color(1.6cm)=(wl540);  
  color(1.7cm)=(wl550);
  color(1.8cm)=(wl560);
  color(1.9cm)=(wl570);
  color(2cm)=(wl580);
  color(2.1cm)=(wl590);
  color(2.2cm)=(wl600);
  color(2.3cm)=(wl610);
  color(2.4cm)=(wl620);
  color(2.5cm)=(wl630);
  color(2.6cm)=(wl640);
  color(2.7cm)=(wl650);
  color(2.8cm)=(wl660);
  color(2.9cm)=(wl670);
  color(3cm)=(wl680);
  color(3.1cm)=(wl690);
  color(3.2cm)=(wl700);
 color(3.7cm)=(wl750);
  color(4cm)=(wl780)
}
\begin{document}
\title{A Flexible \texorpdfstring{$n/2$}{n/2} Adversary Node Resistant and Halting Recoverable Blockchain Sharding Protocol}
\author[1]{Yibin Xu*}
\address[1]{School of Computer Science and Informatics, Cardiff University, Cardiff, UK}
\author[2]{Yangyu Huang}
\address[2]{Guilin University of Electronic and Technology, Guilin, China}
\author[1]{Jianhua Shao}
\author[1]{George Theodorakopoulos}
\corres{Yibin Xu, School of Computer Science and Informatics, Cardiff University, Cardiff, UK, 
\email{work@xuyibin.top}}
\abstract[Summary]{
Blockchain sharding is a promising approach to solving the dilemma between decentralisation and high performance (transaction throughput) for blockchain. The main challenge of Blockchain sharding systems is how to reach a decision on a statement among a sub-group (shard) of people while ensuring the whole population recognises this statement. Namely, the challenge is to prevent an adversary who does not have the majority of nodes globally but have the majority of nodes inside a shard. Most Blockchain sharding approaches can only reach a correct consensus inside a shard with at most $n/3$ evil nodes in a $n$ node system. There is a blockchain sharding approach which can prevent an incorrect decision to be reached when the adversary does not have $n/2$ nodes globally. However, the system can be stopped from reaching consensus (become deadlocked) if the adversary controls a smaller number of nodes.

In this paper, we present an improved Blockchain sharding approach that can withstand $n/2$ adversarial nodes and recover from deadlocks. The recovery is made by dynamically adjusting the number of shards and the shard size. A performance analysis suggests our approach has a high performance (transaction throughput) while requiring little bandwidth for synchronisation.
}

\maketitle
{\fontfamily{lmr}\selectfont
\section{Introduction}
During the thrive of Cryptocurrency in the past ten years, researchers presented various kinds of Distributed ledgers, e.g., Permissionless Blockchain \cite{nakamoto2008bitcoin,xu2018mwpow,xu2019mwpow}, Directed Acyclic Graph (DAG) \cite{popov2016tangle,benvcic2018distributed,sompolinsky2016spectre}. Many are working on extending the usage of Permissionless blockchain to power Decentralised Autonomous Organizations (DAOs) \cite{norta2015creation}, or Decentralised Autonomous Companies (DAC) \cite{barber1999decentralised}, which gather the anonymous resources throughout the network to collectively perform tests without any centralised manager and the result integrity is guaranteed. For example, by using a scalable blockchain as the foundation technology, a data grid or distributed database can outsource the jobs to the users. Many $IoT$ devices (e,g. Amazon assistants, Google home services) on the one hand, live in the centre of people's privacy (e,g. home, office), while on the other hand they require a significant amount of data being sent to the cloud for analysis. Blockchain may help $IoT$ devices  function in a decentralised manner following a well-defined protocol (smart contract), while reaping the benefits of big data and alleviating the concern over privacy or even espionage.

However, it has been a long-standing question how to open the membership in a distributed system while maintaining the performance, integrity, and correctness of the distributed job results \cite{ishibuchi1992distributed}. Traditional blockchains can only process a limited number of transactions per second, and the decentralisation will be compromised if the number of transactions per second is increased. Adding more workload to existing devices in a network would force the resource-constrained devices which are already exhausted to download and verify updates happening throughout the network continually. Of course, the participants can ease the work burden and increase the reward rate by indirectly mining through mining pools where the minimal computation power is gathered and used collectively by one representative. However, the mining pool participants cannot make a judgment on the way their computation power is used, and as a result the system becomes more centralised, which we would like to avoid. Various approaches have been explored in hopes of solving the blockchain's scalability problem, out of these approaches \cite {pass2017hybrid,popov2016tangle,sompolinsky2016spectre,burchert2018scalable,decker2015fast,kokoris2018omniledger,Xu_2019,xu2018section}, Blockchain sharding \cite {kokoris2018omniledger} is a promising one.

Blockchain sharding is a method that splits the transactions amongst all the participating nodes. By allowing multiple node committees to process incoming transactions in parallel, a sharding-based blockchain protocol can increase its throughput with the increase of the number of participants joining the network. Blockchain sharding has also been used to reduce the storage requirement for non-sharding blockchains \cite{xu2020segment}, which helps the blockchain to be implemented in IoT devices that are lacking storage space. Financial models \cite{xu2020anchoring} can be built into the blockchain sharding approach to link the digital labour and the market behaviour with the changes in pay and service prices. Blockchain sharding can also increase the security of blockchain systems by strengthening decentralisation. A recent analysis \cite{mariem2020all} shows more than $36\%$ of bitcoin nodes are now hosted on only five primary cloud services: This phenomenon can raise a high-security concern, as the cloud services can shut down the bitcoin network suddenly. By employing blockchain sharding,  many more users can host blockchain nodes in resource-constrained devices with less processing power than mining-specialised devices and participate as independent miners in blockchain's mining game. 

An important aim when designing a blockchain sharding architecture is to reduce the chance for many adversary nodes to be assigned into the same sub-network. Situations in which the adversary does not hold the majority in the whole network but dominates a sub-network should be strictly prevented to safeguard the entire blockchain system. Elastico \cite{luu2016secure} is known as the first sharding-based consensus protocol designed for blockchain. Elastico has several drawbacks:  firstly, all nodes need to reset their identities after every iteration, and all the committees (shards) are rebuilt. Secondly, the latency increases linearly with the increase of the network size because it requires much more time to fill up all the shards by solving enough PoWs. Thirdly, the process of the committee selection can be misled by the adversary: the adversary can pre-compute PoW puzzles. Fourthly, the probability of failure is primarily increased because a small-sized committee (of around 100 members) is required to restrict the running Practical adversary Fault Tolerance (PBFT) \cite{su2015premature} in each committee. Elastico has been proved to be insecure in practice \cite{kokoris2018omniledger}, and the failure probability can be as high as $0.97$ after merely six iterations. Every block must be broadcast to all parties, though Elastico allows each participant only to verify a subset of transactions. OmniLedger \cite{kokoris2018omniledger} improves on Elastico, but it can only tolerate $n/4$ adversary nodes \cite{zamani2018rapidchain,kokoris2018omniledger}; its security is safeguarded by a large number of participants inside a shard. Due to the limited number of shards, the ceiling of transaction throughput is still relatively low overall. As there are multiple levels of committees, several elections are needed for the system to form this contribution which slows down the system. RSCoin \cite{danezis2015centrally} is a sharding-based protocol for helping centrally-banked cryptocurrencies to scale. It is an approach to combine a distributed network with a centralised monetary supply bringing transparency to the traditional centralised banking systems. However, it relies on a trusted source. RSCoin is not adversary fault-tolerant because each shard runs on a two-phase committee protocol. RapidChain \cite{zamani2018rapidchain} improves the security threshold to $n/3$ adversary node resistant; however, the shard size is still significant. 

We previously proposed a new approach \cite{XUXU} that classifies the nodes into different classes and maintains the number of nodes of different classes in every shard. This approach raises the security level of Blockchain sharding to $n/2$ adversary node resistant, and mostly shrinks the shard size and improves the throughput. However, every shard must have one and only one node from each class, the global rearrangement of the nodes is often needed when the nodes of some classes of a shard go offline. Many new nodes might be left in a pending state when there are not enough nodes from different classes to form new shards. The shard number does not fit into the data flow so that the shard size is only relevant to the number of nodes in the system instead of the real-time workload. It might result in a situation where many shards have no tests to run while the system is idle, and there are not enough shards when the system is busy. Furthermore, the system might be halted by the adversary who has at least ${(m-T+1)}\times{(n/m)}$ of nodes when the system of \emph{n} nodes rules that the consensus in a shard sized \emph{m} is reached with approval from at least $T$ nodes in that shard. 

In this paper, we present a flexible $n/2$ adversary node resistant blockchain protocol that can adjust the class number, shard number, and shard size based on the workload. Our design in this paper serves as the extension of our previous work \cite{XUXU}, and it solves the halting problem in that work when the adversary has fewer than $n/2$ of nodes. We can achieve the security levels of both the $n/2$ tamper-resistant and $n/2$ halting resistant. Our protocol not only increases the security level compared to other methods at $n/3$ security level at most, but it also increases the number of shards and reduces the size of shards. With an increased shard number (more shards running in parallel, and less workload per shard) and a reduced shard size (the less number of nodes in the shard, less communication among nodes in the shard), we strengthen the decentralisation (universal joinability) and increase the performance of the blockchain.

In the remaining of the paper, we first discuss a traditional blockchain sharding approach in section \ref{1.1}. We then discuss the $n/2$ adversary resistant blockchain sharding approach in section \ref{2}, and why it may be deadlocked when the adversary has ${(m-T+1)}\times{(n/m)}$ of nodes. We then propose a new blockchain sharding approach in section \ref{3}, which can recover from a deadlock (global halting) and achieve the $n/2$ halting resistant security level. In section \ref{4}, we show results from a data and performance analysis, discussing the performance (transaction throughput) and the data requirement for nodes.

\subsection{Blockchain sharding Hypothesis}
\label{1.1}
If people are inside of a forest recording the time when trees fall to the ground, there is no need for everyone in the forest to hear and record every fall to maintain the fairness and integrity of the record. The fact that a tree falls and the time when a tree falls is correct when it is recognised by most people around the tree, assuming these people have not colluded. With a sufficient number of people, if they are assigned randomly and distributed to subareas in the forest and are relocated time by time to avoid the accumulation of adversary power, collusion is hard to happen (expected to occur in hundreds of years).

In particular, this proposal is secure when (1) only people assigned to a subarea of the forest can record the information about this subarea. (2) No person can control or predict which subarea they will be assigned to. (3) The assignment follows a globally recognised rule, not by the arbitrary willing of some specific group of superior people. (4) People are periodically reassigned. (5) The number of people in every shard is a sufficient size.

When people assume there is a sufficient number of people acting honestly; people would only need to check what is the typically recognised time of falling for a tree of their interest from the subarea where this tree belongs. It is not necessary for themselves to hear the falling. In this regard, people do not need to have excellent hearing ability when the forest is dense. Instead, they only need to focus on monitoring the subarea in which they are assigned to.

The challenges in this model are as follows: (1) How to distribute people to subareas in a decentralised and unpredictable way? (2) How can people determine if a record of a subarea is made by people assigned to that area?  (3) Without monitoring what happened in a subarea, how can people outside know if the majority of the population in that area support a record or not? (4) How large is a population and how many subareas does the forest need to make a "collusion" unlikely to happen?

\subsection{Failure Probability}
Assuming there exist methods to solve all or some of these challenges above, we calculate the probability of a "collusion" to happen. The probability of obtaining no less than \emph{X} adversary nodes when randomly picking a shard sized \emph{m} (\emph{m} is the number of nodes inside the shard) can be calculated by the cumulative hypergeometric distribution function without replacement from a population of \emph{n} nodes. Let \emph{X} denote the random variable corresponding to the number of adversary nodes in the sampled shard. The failure probability for one shard is at most:
\begin{equation}
    Pr[X>[m/2]]=\sum^m_{X=[m/2]}\frac{(^t_X)(^{n-t}_{m-X})}{(^n_m)}
    	\label{q1-r1}
\end{equation}
which calculates the probability that no less than \emph{X} nodes are adversary in a shard sized \emph{m} and the adversary has \emph{t} number of nodes globally in a system of \emph{n} nodes. Rapidchain \cite{zamani2018rapidchain} is designed with this failure probability.

Figure \ref{fig:img2-r1} shows the maximum probability to fail with $n=2000$ and $m=n/s$ where \emph{s} is the number of shards.
\begin{figure}[htbp!]
\centering
	\begin{tabular}{ll}
   \includegraphics[width=0.5\textwidth]{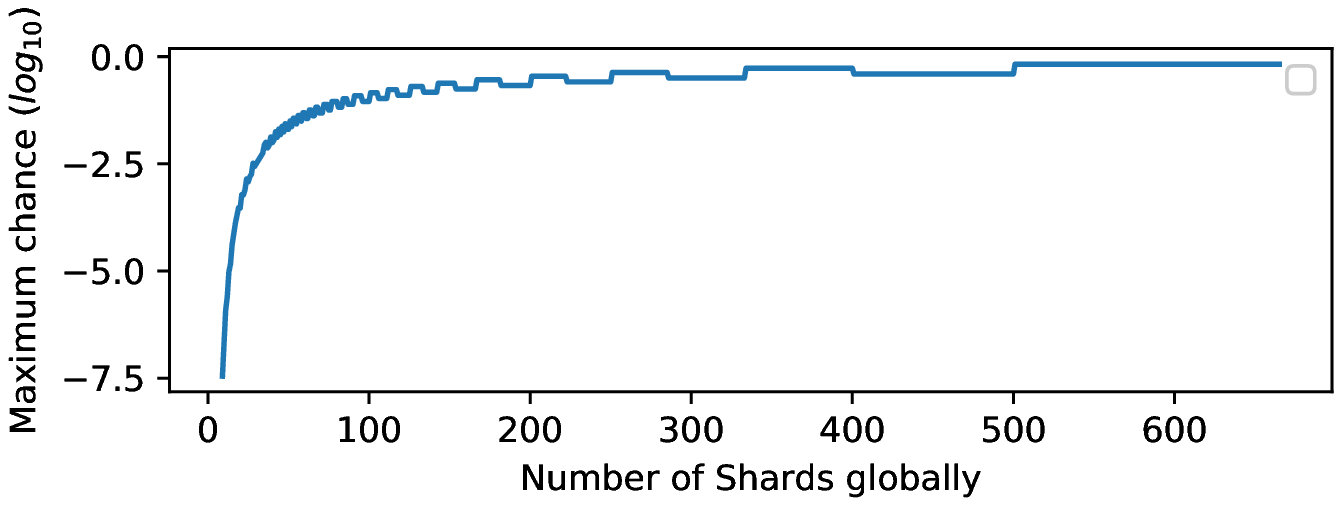}&\includegraphics[width=0.5\textwidth]{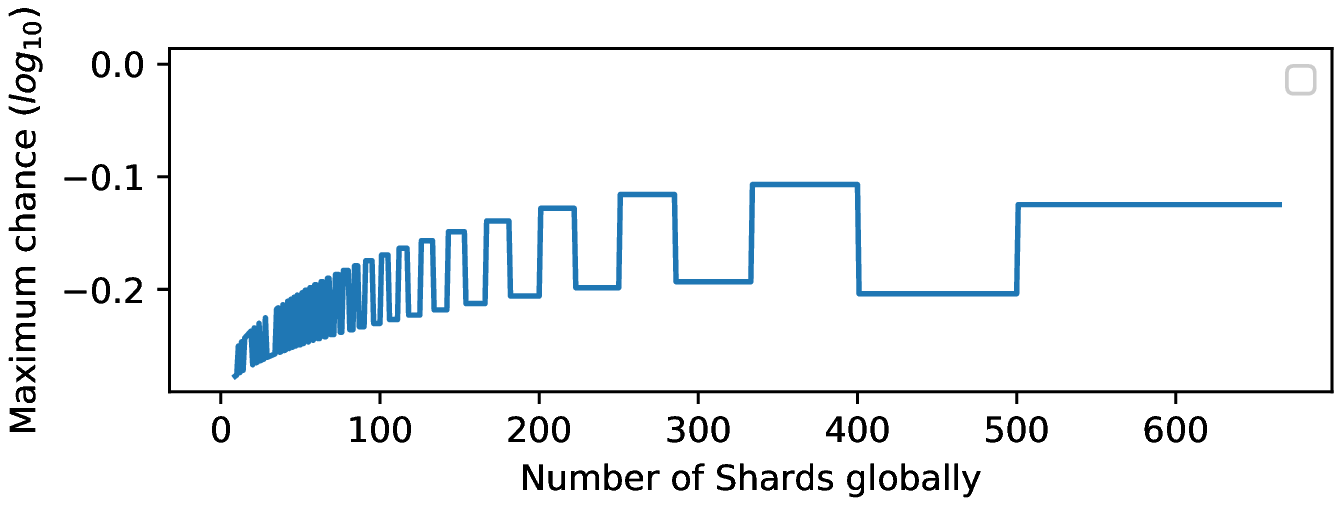}\\
	$t=n/3=666$&$t=n/2=1000$\\
\end{tabular}
	\caption{\fontfamily{lmr}\selectfont The chance to fail when $n=2000$, $t=n/3$, $t=n/2$ and $m=n/s$ where $s$ is the number of shards;}
	\label{fig:img2-r1}
\end{figure}

As can be seen from the result, the system has a very high failure chance when the adversary taken $n/2$ of nodes. The high failure probability is the main reason why most Blockchain sharding approaches can only withstand up to $n/3$ of nodes being evil, and the shard size must be enormous.
\section{\texorpdfstring{$n/2$}{n/2} Blockchain sharding pproach}
\label{2}
\subsection{Hypothesis}
We proposed a $n/2$ adversary node resistant Blockchain sharding approach \cite{XUXU} using a different Hypothesis to build the model. In this hypothesis, nodes are juries inside the courtrooms, and a sentence is made when more than a pre-defined \emph{T} number of people inside the jury sized \emph{m} reached a consensus, $T>\frac{m}{2}$. In order to make the sentences convincing, people inside the jury come from different occupations; united they represent the whole of society. Assuming the rule says a jury should have five people, a teacher, a social worker, a doctor, a businessman, and a police officer. There are five teachers, five social workers, five doctors, five business-people and five police officers for five juries to run in parallel. The \emph{Jury} hypothesis is different from the \emph{Forest} hypothesis because if the adversary controls two doctors, they cannot live inside the same jury. However, they can be inside the same sub-area of the forest when recording the time of the tree falling. 

The challenge of the \emph{Jury} hypothesis is (1) How to give every node a different occupation; and (2) How to make the nodes equally divided into different occupations.
\subsection{Membership management}
Let there be a shard acting as the court office which in charge of Jury schedule arrangement. We refer a pending person as the person who reported to the court office but has not been assigned to a court. Pending people can claim their occupations when reporting to the court office, and they cannot change the occupations once they are assigned into a court. The court office periodically publishes the number of pending people in every occupation. Line up the number of pending people in every occupation in ascending order of the time when they claimed an occupation. Every time there comes \emph{F} pending people in every occupation, the front \emph{F} people in every occupation is added to the courts. In the meantime, \emph{F} more courts are formed, and all the jurors are reassigned into different courts. When assigning people, the court office can also be seen as a court. Namely, every time the juror membership is adjusted, the people inside the court office is also reassigned to other courts. If a pending person changes its occupation, it will go to the tail of the other pending queue, losing the position in the original pending queue. Thus, we expect new participants to check the pending queue of every occupation and choose an unpopular occupation to get into the courts quicker. Otherwise, they will be kept in pending status longer until other people fill in unpopular occupations. As new participants line in the shorter queues, the number of people in every occupation is automatically close to each other (tend to be equal in the long run). The person, regardless if they are inside a court or in the waiting queue, should work (generate \emph{PoW}s in the case of blockchain) in every fixed time window so that the adversary who has half of the overall energy can only have half of the people in the system. 

Table \ref{table:r2},\ref{table:r3},\ref{table:r4},\ref{table:r5} and \ref{table:r6} show the procedure of adding people into the system. In this procedure, we rule that whenever there are at least four pending people in every occupation, the adding starts ($F=4$). Table \ref{table:r2} shows the pending queue published by the court office in moment one. Table \ref{table:r3} shows the pending queue when the adding conditions are met after moment one and before moment two. Table \ref{table:r4} shows the pending queue in moment two. Table \ref{table:r5} shows the people in the court system at in moment one. Table \ref{table:r6} shows the people in the court system at the moment two where some new people are added.
\begin{table}[ht!]
\footnotesize
	\caption{\fontfamily{lmr}\selectfont The pending queue published by the court office at moment 1. Letters in red colors are the pending people. Following the alphabetical order, the letters also reflect the time when people report to the office. }
\centering
\fontfamily{lmr}\selectfont
\begin{tabular}{llllll}
Occupation &I                                                                                             & II                                                                                             & III                                              & IV                                               & V                                                \\
                        &                                                  &                                                  &                                                  &                                                  &                                                  \\
                        & \cellcolor[HTML]{FE0000}{\color[HTML]{FFFFFF} A} &                                                  &                                                  &                                                  &                                                  \\
                        & \cellcolor[HTML]{FE0000}{\color[HTML]{FFFFFF} B} &                                                  &                                                  &                                                  &                                                  \\
                        & \cellcolor[HTML]{FE0000}{\color[HTML]{FFFFFF} C} & \cellcolor[HTML]{FE0000}{\color[HTML]{FFFFFF} E} &                                                  &                                                  &                                                  \\
                        & \cellcolor[HTML]{FE0000}{\color[HTML]{FFFFFF} D} & \cellcolor[HTML]{FE0000}{\color[HTML]{FFFFFF} F} & \cellcolor[HTML]{FE0000}{\color[HTML]{FFFFFF} G} & \cellcolor[HTML]{FE0000}{\color[HTML]{FFFFFF} H} & \cellcolor[HTML]{FE0000}{\color[HTML]{FFFFFF} I} \\
Number of Pending people & 4                                                & 2                                                & 1                                                & 1                                                & 1                                               
\end{tabular}

	\label{table:r2}
\end{table}

\begin{table}[ht!]
\footnotesize
\centering
	\caption{\fontfamily{lmr}\selectfont The pending queue after moment 1 before moment 2. letters in the blue colour represent the people who reported to the office after moment 1 before moment 2. Some nodes will be added to the system afterward because the minimum length of the queues is equal to four (pre-defined adding parameters)}
\fontfamily{lmr}\selectfont
\begin{tabular}{llllll}
Occupation                                               & I                                                & II & III                                              & IV                                               & V                                                                     \\
                                                         &                                                  &                                                  & \cellcolor[HTML]{3531FF}{\color[HTML]{FFFFFF} U} &                                                  & \cellcolor[HTML]{3531FF}{\color[HTML]{FFFFFF} V}                      \\ \cline{2-6} 
\multicolumn{1}{l|}{}                                    & \cellcolor[HTML]{FD6864}{\color[HTML]{FFFFFF} A} & \cellcolor[HTML]{6665CD}{\color[HTML]{FFFFFF} Q} & \cellcolor[HTML]{6665CD}{\color[HTML]{FFFFFF} R} & \cellcolor[HTML]{6665CD}{\color[HTML]{FFFFFF} S} & \multicolumn{1}{l|}{\cellcolor[HTML]{6665CD}{\color[HTML]{FFFFFF} T}}  \\
\multicolumn{1}{l|}{Add to court system -\textgreater{}} & \cellcolor[HTML]{FD6864}{\color[HTML]{FFFFFF} B} & \cellcolor[HTML]{6665CD}{\color[HTML]{FFFFFF} M} & \cellcolor[HTML]{6665CD}{\color[HTML]{FFFFFF} N} & \cellcolor[HTML]{6665CD}{\color[HTML]{FFFFFF} O} & \multicolumn{1}{l|}{\cellcolor[HTML]{6665CD}{\color[HTML]{FFFFFF} P}} \\
\multicolumn{1}{l|}{}                                    & \cellcolor[HTML]{FD6864}{\color[HTML]{FFFFFF} C} & \cellcolor[HTML]{FD6864}{\color[HTML]{FFFFFF} E} & \cellcolor[HTML]{6665CD}{\color[HTML]{FFFFFF} J} & \cellcolor[HTML]{6665CD}{\color[HTML]{FFFFFF} K} & \multicolumn{1}{l|}{\cellcolor[HTML]{6665CD}{\color[HTML]{FFFFFF} L}} \\
\multicolumn{1}{l|}{}                                    & \cellcolor[HTML]{FD6864}{\color[HTML]{FFFFFF} D} & \cellcolor[HTML]{FD6864}{\color[HTML]{FFFFFF} F} & \cellcolor[HTML]{FD6864}{\color[HTML]{FFFFFF} G} & \cellcolor[HTML]{FD6864}{\color[HTML]{FFFFFF} H} & \multicolumn{1}{l|}{\cellcolor[HTML]{FD6864}{\color[HTML]{FFFFFF} I}} \\ \cline{2-6} 
Number of Pending people                                  & 4                                                & 4                                                & 5                                                & 4                                                & 5                                                                    
\end{tabular}

	\label{table:r3}
\end{table}

\begin{table}[ht!]
\footnotesize
\centering
	\caption{\fontfamily{lmr}\selectfont The pending queue published by the court office at moment 2. Selected people in Table \ref{table:r3} has been assigned to the system.}
\fontfamily{lmr}\selectfont
\begin{tabular}{llllll}
Occupation              & I & II & III                      & IV & V                         \\
                        &   &   &                           &    &                           \\
                        &   &   &                           &    &                           \\
                        &   &   &                           &    &                           \\
                        &   &   &                           &    &                           \\
                        &   &   &  \cellcolor[HTML]{FE0000}{\color[HTML]{FFFFFF} U} &    &  \cellcolor[HTML]{FE0000}{\color[HTML]{FFFFFF} V} \\
Number of Pending people & 0 & 0 & 1                         & 0  & 1                        
\end{tabular}
	\label{table:r4}
\end{table}
\begin{table}[ht!]
\footnotesize
\centering
\fontfamily{lmr}\selectfont
	\caption{\fontfamily{lmr}\selectfont People in the court system at moment 1. There is only one jury running.}
\begin{tabular}{llllll}
\fontfamily{lmr}\selectfont
\diagbox{Ocp}{Court}     & 1                                                 &                                                 &                                                 &                                                 &                                                 \\
Occupation I   & \cellcolor[HTML]{FFFFFF}{\color[HTML]{000000} !}  & \cellcolor[HTML]{FFFFFF}{\color[HTML]{000000} } & \cellcolor[HTML]{FFFFFF}{\color[HTML]{000000} } & \cellcolor[HTML]{FFFFFF}{\color[HTML]{000000} } & \cellcolor[HTML]{FFFFFF}{\color[HTML]{000000} } \\
Occupation II  & \cellcolor[HTML]{FFFFFF}{\color[HTML]{000000} @}  & \cellcolor[HTML]{FFFFFF}{\color[HTML]{000000} } & \cellcolor[HTML]{FFFFFF}{\color[HTML]{000000} } & \cellcolor[HTML]{FFFFFF}{\color[HTML]{000000} } & \cellcolor[HTML]{FFFFFF}{\color[HTML]{000000} } \\
Occupation III & \cellcolor[HTML]{FFFFFF}{\color[HTML]{000000} \#} & \cellcolor[HTML]{FFFFFF}{\color[HTML]{000000} } & \cellcolor[HTML]{FFFFFF}{\color[HTML]{000000} } & \cellcolor[HTML]{FFFFFF}{\color[HTML]{000000} } & \cellcolor[HTML]{FFFFFF}{\color[HTML]{000000} } \\
Occupation IV  & \cellcolor[HTML]{FFFFFF}{\color[HTML]{000000} \$} & \cellcolor[HTML]{FFFFFF}{\color[HTML]{000000} } & \cellcolor[HTML]{FFFFFF}{\color[HTML]{000000} } & \cellcolor[HTML]{FFFFFF}{\color[HTML]{000000} } & \cellcolor[HTML]{FFFFFF}{\color[HTML]{000000} } \\
Occupation V   & \cellcolor[HTML]{FFFFFF}{\color[HTML]{000000} *}  & \cellcolor[HTML]{FFFFFF}{\color[HTML]{000000} } & \cellcolor[HTML]{FFFFFF}{\color[HTML]{000000} } & \cellcolor[HTML]{FFFFFF}{\color[HTML]{000000} } & \cellcolor[HTML]{FFFFFF}{\color[HTML]{000000} }
\end{tabular}
	\label{table:r5}
\end{table}
\begin{table}[ht!]
\footnotesize
	\caption{\fontfamily{lmr}\selectfont People in the court system at moment 2. All membership is adjusted with four more juries formed.}
\fontfamily{lmr}\selectfont
\centering
\begin{tabular}{llllll}
\diagbox{Ocp}{Court}     & 1                                                & 2                                                & 3                                                 & 4                                                & 5                                                 \\
Occupation I   & \cellcolor[HTML]{FFFFFF}{\color[HTML]{000000} A} & \cellcolor[HTML]{FFFFFF}{\color[HTML]{000000} C} & \cellcolor[HTML]{FFFFFF}{\color[HTML]{000000} B}  & \cellcolor[HTML]{FFFFFF}{\color[HTML]{000000} D} & \cellcolor[HTML]{FFFFFF}{\color[HTML]{000000} !}  \\
Occupation II  & \cellcolor[HTML]{FFFFFF}{\color[HTML]{000000} @} & \cellcolor[HTML]{FFFFFF}{\color[HTML]{000000} Q} & \cellcolor[HTML]{FFFFFF}{\color[HTML]{000000} M}  & \cellcolor[HTML]{FFFFFF}{\color[HTML]{000000} F} & \cellcolor[HTML]{FFFFFF}{\color[HTML]{000000} E}  \\
Occupation III & \cellcolor[HTML]{FFFFFF}{\color[HTML]{000000} R} & \cellcolor[HTML]{FFFFFF}{\color[HTML]{000000} N} & \cellcolor[HTML]{FFFFFF}{\color[HTML]{000000} \#} & \cellcolor[HTML]{FFFFFF}{\color[HTML]{000000} J} & \cellcolor[HTML]{FFFFFF}{\color[HTML]{000000} G}  \\
Occupation IV  & \cellcolor[HTML]{FFFFFF}{\color[HTML]{000000} S} & \cellcolor[HTML]{FFFFFF}{\color[HTML]{000000} O} & \cellcolor[HTML]{FFFFFF}{\color[HTML]{000000} H}  & \cellcolor[HTML]{FFFFFF}{\color[HTML]{000000} K} & \cellcolor[HTML]{FFFFFF}{\color[HTML]{000000} \$} \\
Occupation V   & \cellcolor[HTML]{FFFFFF}{\color[HTML]{000000} T} & \cellcolor[HTML]{FFFFFF}{\color[HTML]{000000} *} & \cellcolor[HTML]{FFFFFF}{\color[HTML]{000000} P}  & \cellcolor[HTML]{FFFFFF}{\color[HTML]{000000} I} & \cellcolor[HTML]{FFFFFF}{\color[HTML]{000000} L} 
\end{tabular}
	\label{table:r6}
\end{table}

As can be seen from the adding procedure, if the adversary is not in a very long pending queue, there is no gain for the adversary to change the occupation once it reports to the court office. If it does so, it goes to the tail of another queue, leaving its original place to others. With a sufficient number of participants, only an insignificant number of people will be left outside the system eventually.
\subsection{Failure Probability}
Table \ref {fig:img3-r1} shows a court schedule table for five courts run in parallel with the jury sized five (five people in different occupations). \emph{A} refers to an adversary person. \emph{H} refers to an honest person.
\begin{table}[ht!]
\footnotesize
\caption{\fontfamily{lmr}\selectfont Court Jury Schedule}
\centering
\fontfamily{lmr}\selectfont
\begin{tabular}{cccccc}
\diagbox{Ocp}{Court}&0&1&2&3&4\\
Occupation I&A&A&A&A&A\\
Occupation II&H&A&H&A&H\\
Occupation III&A&H&A&H&A\\
Occupation IV&H&A&H&H&A\\
Occupation V&H&H&H&H&A\\
\end{tabular}
\label{fig:img3-r1}
\end{table}

For a \emph{s} number of courts to be held in parallel, there is an \emph{s} number of people in each occupation. To manipulate a sentence in of a court, the adversary must gain control of at least \emph{T} people inside this court. Because the adversary not having more than $\frac{n}{2}$ of people globally and $T>\frac{m}{2}$, so \begin{equation}
    T\times s > t
\end{equation} where \emph{t} is the number of adversary persons. If the adversary has $A_i$ number of the person in occupation \emph{i}; then the chance for the adversary to secure a manipulated sentence is (assuming it place all the adversary person in the front \emph{T} occupations): 
\begin{equation}
 Pr[T]=\prod_{i=1}^{T}\frac{A_i}{s}
\end{equation}

To derive the maximised value of $Pr[T]$, we want $\prod_{i=1}^{T}A_i$ to be maximised because \emph{s} is the same number (every occupation has \emph{s} people inside the system). If the adversary has \emph{t} number of people inside the system (Court Jury Schedule), then \begin{equation}
    t=\sum_{i=1}^m A_i
\end{equation} To let the value of $\prod_{i=1}^{T}A_i$ maximised, we consider \begin{align}
    A_i &= \lceil(t/T)\rceil, \quad i \in [1,t \bmod T]\\
    A_i &= \lfloor(t/T)\rfloor, \quad i \in (t \bmod T, T]
\end{align}
This scenario is maximised because given any positive Integer $X$, 
\begin{equation}
    X\times X > (X-1)\times (X+1)=X\times X-1
\end{equation}

Thus, \begin{equation}
    Pr[T]_{max}\approx(\frac{t}{T\times s})^T
\end{equation}

Though the adversary cannot manipulate a sentence when it does not have \emph{T} people inside a shard, it can halt a sentence to be reached when it has \emph{m-T+1} number of the nodes in a shard. This sentence cannot then be made until the next court (the group of juries is re-selected). Thus, to make the system function more smoothly, we want $T\approx[m/2]$ whilst meeting the security threshold (e.g., $10^{-6}$ failure chance). Figure \ref{fig:img4-r1} shows the maximum failure chance with different \emph{s},$n=s\times m=2000$, $T=0.7\times m$ and $t=1000$ ($1/2$ fraction of the overall population).\begin{figure}[htbp!]\centering
	\begin{tabular}{ll}
   \includegraphics[width=0.5\textwidth]{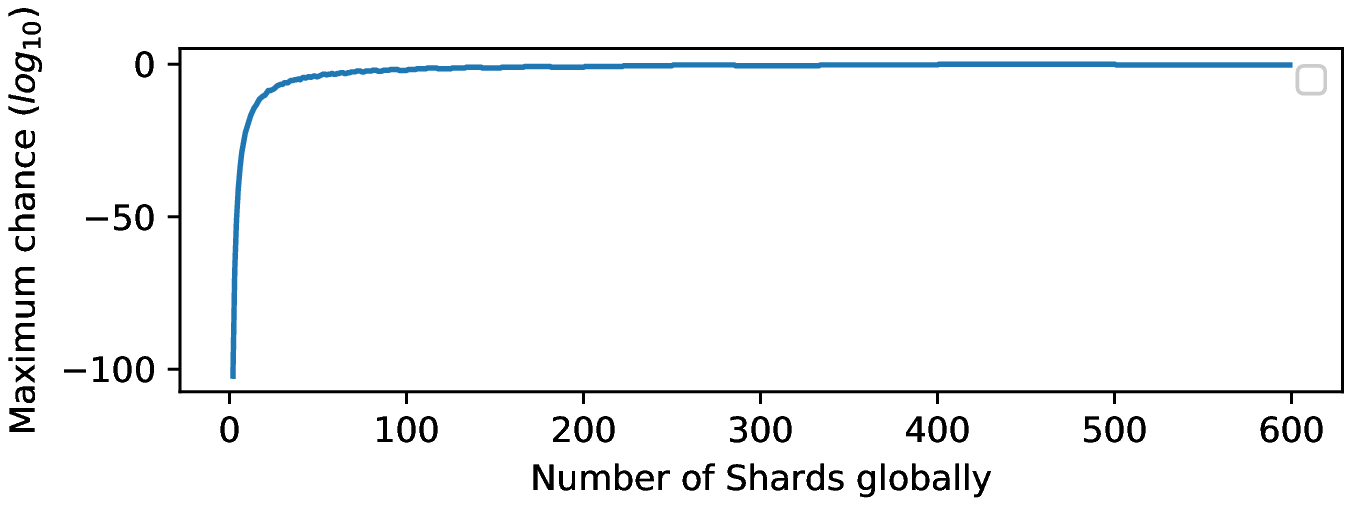}&\includegraphics[width=0.5\textwidth]{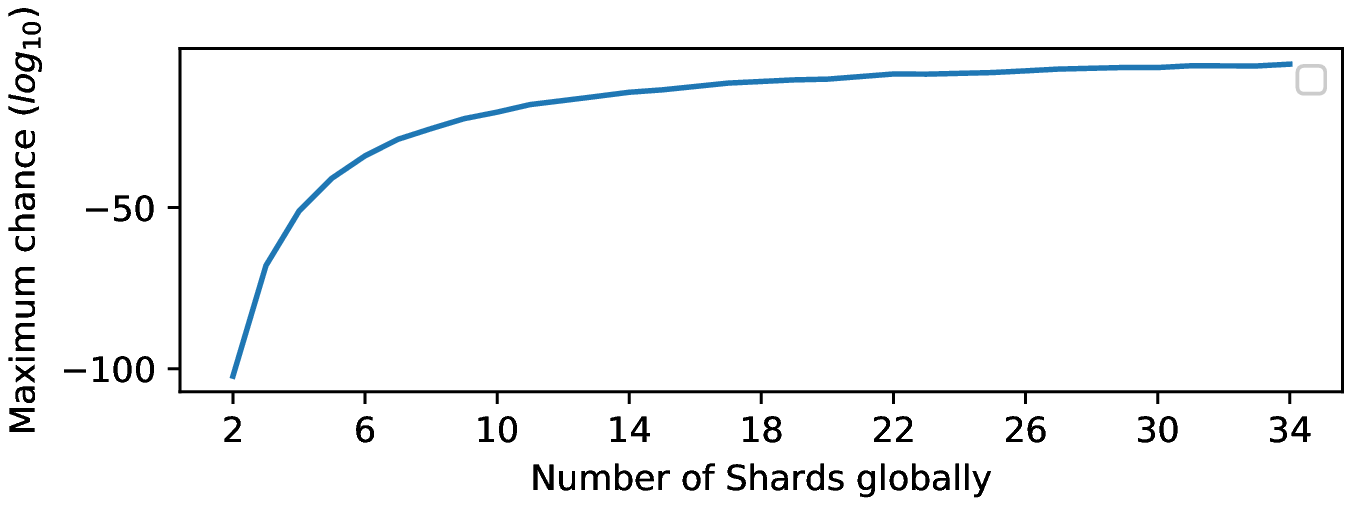}\\
{$s \in [2,600], t=1000=n/2, T=0.7\times m$}&
	{$s \in [2,34], t=1000=n/2, T=0.7\times m$}\\
\end{tabular}
	\caption{\fontfamily{lmr}\selectfont The chance to fail with different \emph{s} when $n=2000$ and $m=n/s$ where \emph{s} is the number of shards;}
	\label{fig:img4-r1}
\end{figure}

Comparing Figure \ref{fig:img2-r1} and Figure \ref{fig:img4-r1}, the $Jury$ hypothesis largely outperforming the $forest$ hypothesis even when there are $n/2$ of evil nodes in the $Jury$ hypothesis and only $n/3$ of evil nodes in the $forest$ hypothesis. 
When $n=2000$, if maintaining a $10^{-6}$ failure chance, $Jury$ ($n/2$ evil nodes) can have 33 shards at the same time, while $forest$ ($n/3$ evil nodes) can only have 10 shards.

\subsection{Drawbacks}
The system as a whole is halted when the adversary took $s\times(m-T+1)$ of nodes, and it places all of these nodes into the same $m-T+1$ occupations. In this scenario, it is guaranteed that the adversary has $m-T+1$ of nodes inside every shard. Table \ref{fig:img444-r1} shows a example of a system halting, where $m=5$ and $T=4$, the adversary took $m-T+1$ number of nodes in every shard.

\begin{table}[ht!]
\footnotesize
\caption {\fontfamily{lmr}\selectfont A halted scenerio}
\fontfamily{lmr}\selectfont
\centering
\begin{tabular}{cccccc}
\diagbox{Ocp}{Court}&0&1&2&3&4\\
Occupation I&A&A&A&A&A\\
Occupation II&A&A&A&A&A\\
Occupation III&H&H&H&H&H\\
Occupation IV&H&H&H&H&H\\
Occupation V&H&H&H&H&H\\
\end{tabular}
\label{fig:img444-r1}
\end{table}

Because the court office (the shard which deals with the node membership) is also stopped from generating further blocks that can be approved by $T$ people inside it, the pending nodes cannot be added to the system. Thus, though the $n/2$ Blockchain sharding method secured the system from tampering when the adversary is below $n/2$, the system can stop functioning anymore when $s\times(m-T+1)$ of nodes are evil.
\section {Flexible \texorpdfstring{$n/2$}{n/2} Blockchain sharding approach}
\label{3}
In this section, we introduce a flexible $n/2$ Blockchain sharding approach that can dynamically adjust $m$ in shards to defeat the adversary. It takes $n/2$ nodes for the adversary to tamper a record or halt the system eventually.  
\subsection{Hypothesis}
We use a new hypothesis named \emph{colour} Hypothesis. It is ruled that every node should claim a colour from the colour spectrum when joining in the system — every node inside a shard is categorised to its closet \emph{base colour}. If there is a \emph{m} number of nodes inside a shard, then there is a \emph{m} number of \emph{base\ colours} globally which together represent the colour spectrum as a whole. The same as the \emph{Jury} Hypothesis, a consensus of a shard should be approved by at least a pre-defined \emph{T} people (\emph{T} must larger than $0.5 \times m$) inside this shard. Figure \ref{fig:img5-r1} shows the example of the \emph{base\ colour}. Every \emph{base colour} should have the same number of nodes inside. This hypothesis is different from the \emph{jury} Hypothesis, the \emph{jury} Hypothesis requires a fixed number of occupations, however this hypothesis can regroup a \emph{colour} to a different \emph{base\ colour} base on the change of the number of \emph{base\ colours}.

The challenge of \emph{colour} Hypothesis is to make the number of nodes in every categorisation the same as each other, especially with the rapid changes of the number of \emph{base\ colour}.
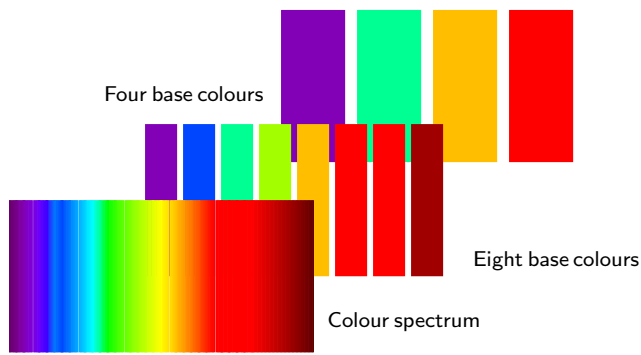
\begin{figure}[htbp]\centering
\begin{tikzpicture}
\foreach[evaluate={\x=\wl/100}]\wl in {400,500,600,700}
  \draw[line width=24pt,wl\wl] (\x-2,1.5)--(\x-2,3.5);
\foreach[evaluate={\x=\wl/100}]\wl in {400,450,...,750}
  \draw[line width=12pt,wl\wl] (\x-4,0)--(\x-4,2);

\node[inner sep=3pt]{\pgfuseshading{spectrum}};
\node at (3.2,-0.6){{$Colour\ spectrum$}} ;
\node at (5.2,0.2){{$Eight\ base\ colours$}} ;
\node at (0.3,2.4){{$Four\ base\ colours$}} ;
\end{tikzpicture}
\caption{\fontfamily{lmr}\selectfont The colour spectrum and $base\ colours$}
\label {fig:img5-r1}
\end{figure}
\subsection{Failure Probability}
Assuming the adversary has a $t=\frac{n}{2}-1$ number of nodes. The chance for the adversary to take control of a shard in a system which has \emph{n} nodes, $m$ \emph{base\ colours} and $s$ number of shards ($s=n/m$) is:
\begin{equation}
    Pr[T]_{Max}={(\frac{t}{T\times s})}^{T}={(\frac{t/T}{n/m})}^{T}\approx{(\frac{m}{2\times T})}^{T}
\end{equation}

As can be seen from Figure \ref{fig:img6-r1}, \emph{T} can be adjusted by the number of \emph{m} when maintaining a fixed threshold failure chance. When \emph{m} is over $800$, $T/m$ is very close to $0.5$ (the adversary needs to take approximately $n/2$ of nodes to halt the system). We aim to dynamically adjust \emph{m} to increase the difficulty for the adversary in deadlocking a shard.
\begin{figure}[htbp]
\centering
   \includegraphics[width=0.5\textwidth]{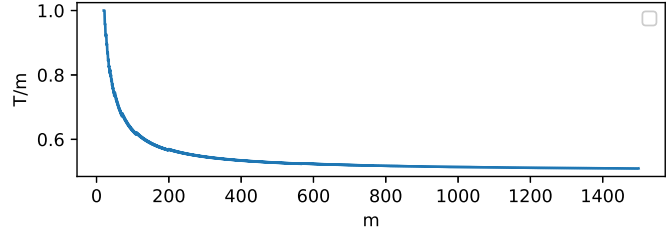}
	\caption{\fontfamily{lmr}\selectfont \emph{T/m} for maintaining a $10^{-6}$ failure chance with different \emph{m}}
	\label{fig:img6-r1}
\end{figure}
\subsection{Model}
We run a committee to deal with the membership issues (the same to the court office in \emph{Jury} Hypothesis). The new nodes report to the committee and wait for the assignment by the committee. The committee's block contains two sections. 
\begin{itemize}
\item Pending node section.The node information of nodes which reported to the committee before but have not yet assigned to a shard.
\item Member node section. The node information of nodes inside each shard.
\end{itemize}

When a node wants to join the system, the node will check the information in the Pending node section and the information in the Member node section. It then selects a colour code that is more likely to be added to the system. It then tells the committee the colour code it chooses and its public identity key. When a node's information is added to the Pending node section, it needs to continuously send one \emph{PoW} to the committee per iteration to maintain its spot in the section until it is assigned. After the node is assigned to a shard, it also needs to present one \emph{PoW} per iteration. In this way, we prevented the node simulation problem: if the adversary has $50\%$ of the overall power, it can only simulate $n/2$ of nodes. The system will add the pending nodes to make the number of shards fulfilling the current workload. There is a number of pending transactions indicated in the block header of every shard. We pre-define a standard workload \emph{K} which is also written in the block header of every shard. When the pending transaction of a shard exceeds \emph{2K} there should be one more shard added to the system. If the pending transaction is below $\frac{K}{2}$, then the system should cut this shard. In this way, a shard has at least \emph{K/2} workload and at most \emph{2K} workload. 

Let there be $m=\frac{n}{s}$ colour categorisations (\emph{base\ colour}) currently. 

Let \begin{equation}
    mt=\frac{n+snp}{st}
\end{equation}
where \emph{st} is the ideal number of shards in the system based on the current workload, \emph{snp} is the number of pending nodes that should be added to the system for the system to achieve \emph{st} number of shards (\emph{snp} can be zero) and \emph{mt} is the number of colour categorisations corresponding to that situation. Find a \emph{snp} number of nodes from the pending nodes that after adding these pending nodes, there can be \emph{st} number of shards with each shard sized \emph{mt}. If such \emph{snp} number of nodes are found, then the shard number is adjusted to \emph{st}, and the selected nodes are added. The deselected nodes remain in the pending section. If there is more than one \emph{snp} possible, then it should select the largest \emph{snp}. 

If a node inside a shard did not show a \emph{PoW} in an iteration, it is labelled as \emph{offline}, and this information is written into the next block header of that shard. All the shards (including the committee) synchronise the block headers of the blocks in other shards, and these blocks should have at least \emph{T} approvals in the relevant shards. Then, the committee updates its Member node section using such information. When a block of a shard is not approved by at least \emph{T} people inside this shard in a fixed time frame (e.g., \emph{10\ minutes}), the block is abandoned. The leader of the next block height will create a new block, and people will vote again for this new block. If the shard cannot reach a consensus in a longer fixed time frame, the shard is halted. All shards know if a shard has halted because they sync the qualified block headers of other shards and the block of the committee. If a shard is halted, the membership of all shards is rearranged (without adjusting the shard number).

If the system is halted (every shard is halted for a fixed time, at the same time), then the nodes use the node information in the committee's latest block to calculate a \emph{st<s} that can make every \emph{base\ colour} of a \emph{st} number of \emph{base\ colours} contains $mt$ number of nodes. If such \emph{snp} is not found, then the shard size is directly reduced to one (all the nodes are inside the same shard). If there exists a \emph{snp} (\emph{snp} can be zero), then the system is rearranged to \emph{st} number of shards. By adjusting the number of colour categorisations, we bring more nodes to shards, so that \emph{T} is respectively reduced. At the extreme, when the shard size is one, the system can withstand the adversary that has a \emph{n/2-1} number of nodes. Especially if it is $s=1$, then there is no colour categorisation restriction (number of nodes inside every categorisation can be different). When the system recovers from the halting problem, the shard number can then be increased base on the data flow again. Figure \ref{fig:img2} shows an illustration of the model.
\begin{figure}
\centering
{

\includegraphics[width=0.55\textwidth]{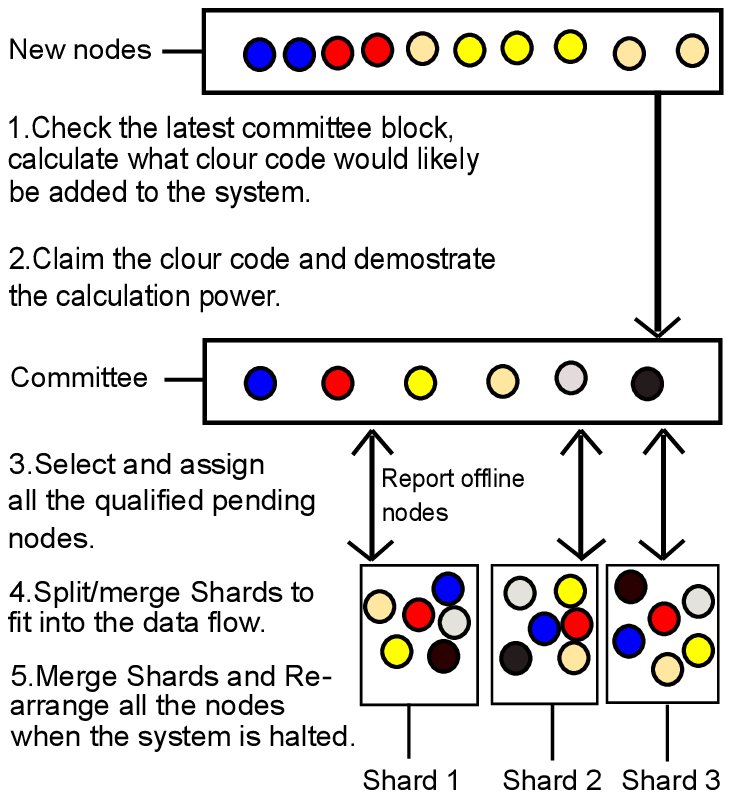}
}
\caption {\fontfamily{lmr}\selectfont The Model.}
\label {fig:img2}
\end{figure}
\subsection{Node assignment}
We refer to the committee as \emph{Shard 0}, and other shards also have their unique shard id. When assigning/rearranging nodes, let $C_{Hash}$ be the block header hash of the latest block (including the signatures of the nodes) that has been approved by at least \emph{T} number of nodes in the \emph{Shard 0}. Let \emph{ID[i][j]} be the public identity key of the node in colour categorisation \emph{i} in the \emph{j} shard. Let '\emph{hash}' be a hash function that returns a $2^{256}$ Integer, calculate a \emph{RID}, where \begin{equation}
RID[i][j]=C_{Hash}\ hash\ ID[i][j]\end{equation} Rank every public key in \emph{ID[i]} by the ascending order of \emph{RID[i]} and then the new assignment will be completed.
\subsection{Pending node selection and node categorisation}
Rank all the nodes in the Pending node section and nodes in the Member node section together into a list by the increasing order of their colour code. If node $x$ is originally from the Pending node section then $D(x)=1$, otherwise $D(x)=0$. Let $F[x][y]$ denote the node assignment scheme for grouping nodes which are selected from the front $x$ nodes in the ranked list; each group should have \emph{y} people inside it. 
\begin{equation}F[x][y]=Max(F[i \in [y...x-y]][y], select (i,x,y))\end{equation} where \emph{select} is a function that takes \emph{y} number of nodes from the node \emph{i} to node \emph{x}. The selection rule is: (1) if $D(x)=0$ then the node \emph{x} must be selected. (2) Node \emph{x} should be selected instead of node \emph{x1} if node \emph{x} has been continuously written in the 'Pending node' section longer than node \emph{x1}. $Max (a, b)$ is a function that: (1) calculates the number of nodes from the pending node section in assignment scheme \emph{a} and \emph{b}; and (2) returns $a$ or $b$ which has a larger number in the step (1).

After the iterations, the assignment scheme in $Max(F[st..n][st])$ is the new node assignment scheme. If such an assignment scheme does not exist, we do not add the pending nodes into the system in that round of the mining game. If the system is halted, then the colour categorisation is automatically re-grouped to $Max(F[ss..n][ss])$, where \emph{ss} is the largest one in \emph{ss<s} that make a $F[ss..n][ss]$ exist. The system then automatically re-arranges people following the rule discussed in the \emph{model} section. In this way, the system will eventually recover from the halt, at the extreme, when $ss=1$.

\subsection{Shard consensus and Global consensus}
Let the block header hash of shard \emph{i} in the block height \emph{BH} be \emph{Blockhash[i][BH]}. $L_i$ is the index number of the random leader of a shard \emph{i} in \emph{BH+1}
\begin{equation}
    L_i=hash(Blockhash[i-5][BH],...,Blockhash[i+5][BH])\ mod\ m
\end{equation}

Specially, 
\begin{equation}
    Blockhash[j][BH]=Blockhash[abs(NS-j)][BH],j \in[-4,0] \cup [NS+1,NS+4]
\end{equation}

If the shard \emph{i} did not reach a consensus on the block height $BH$ then \begin{equation}
    Blockhash[i][BH]=Blockhash[i][BH-1]
\end{equation}

When the leader proposed a block \emph{Bob}, all the nodes inside the shard will sync \emph{Bob} and do the verification. If they approve \emph{Bob}, they will sign \emph{Bob} using their private identity key and attach a \emph{PoW} which fulfils the difficulty they claimed before. If a node does not approve \emph{Bob}, it will not send the signature, but it still needs to send the \emph{PoW}. When a block receives a higher than \emph{T} number of signature as well as the corresponding \emph{PoW}s, then this block (\emph{Bob}) is approved. If a node did not show the \emph{PoW}, then this node is considered \emph{offline} and its info is written to the block header of the next block. Later this node's information will be synchronised by the committee, and this node is erased from the 'Member node' section.

A \emph{detect} then \emph{verify} mechanism is used for the global synchronous. The nodes of a shard will inform other shards when more than \emph{T} nodes approved a block of this shard. Then the nodes of the other shards will download the block header of this block. If a node in this shard is opposing this block, it will inform the other shards. When a conflict is \emph{detected}, the honest nodes in this shard will send the signatures of this block to the other shards for \emph{verification}. Other shards will accept this block if there are at least \emph{T} number of signatures signed by different nodes in this shard. If there is no conflict reported within a given time frame, then other shards will recognise this block as a finally accepted block of this shard. 

All shards can simply verify the signatures received to determine the genuine because all nodes sync the block of the committee; they know all the public keys of nodes. To prevent the adversaries from causing additional verification, the node which sends the global consensus information should sign the data using its private key. In this way, the dishonest node can be labelled as \emph{offline} directly after the verification.

\section{Performance analyses and Data requirement}
\label{4}
In Figure \ref{fig:img84-r1}, we show the minimum percentage of nodes which must be taken by the adversary to halt the entirety of the system with the different number of Shards in two systems of $1500$ nodes and $20000$ nodes. We can observe from the result that if the adversary attempts to halt the system and it has a great number of nodes, the number of shards can be minimal, vise versa. The adjustment of shard number can be done in real time follows the rules of our approach so that we do not need to assume the adversary having $n/2$ of nodes as like calculating the failure probability.

\begin{figure}[ht!]
\centering
	\begin{tabular}{ll}
   \includegraphics[width=0.5\textwidth]{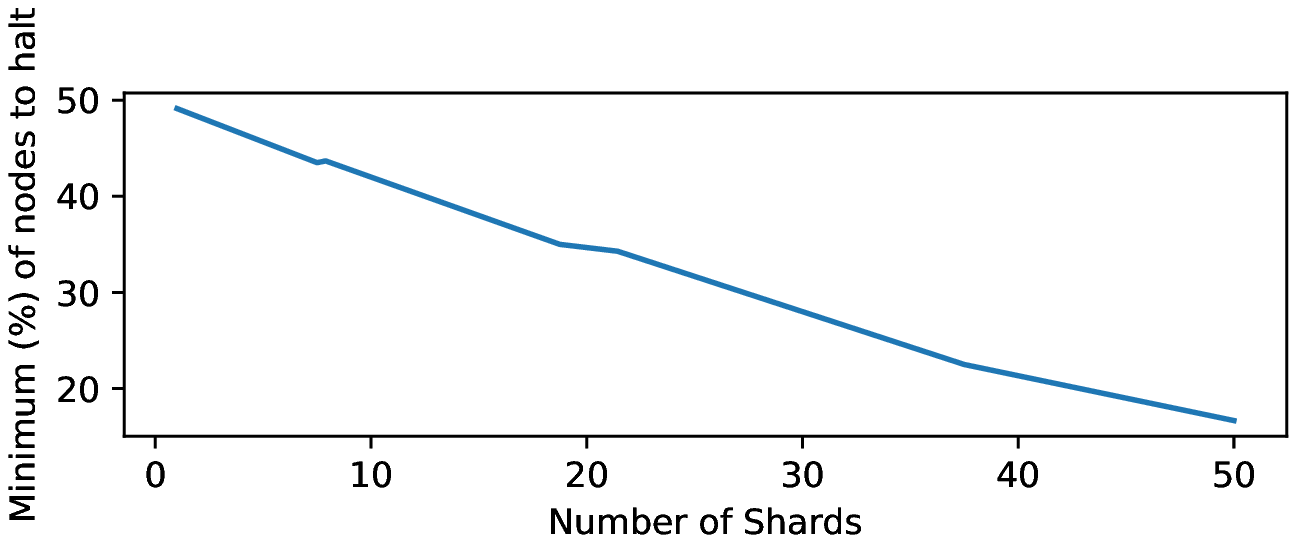}&\includegraphics[width=0.5\textwidth]{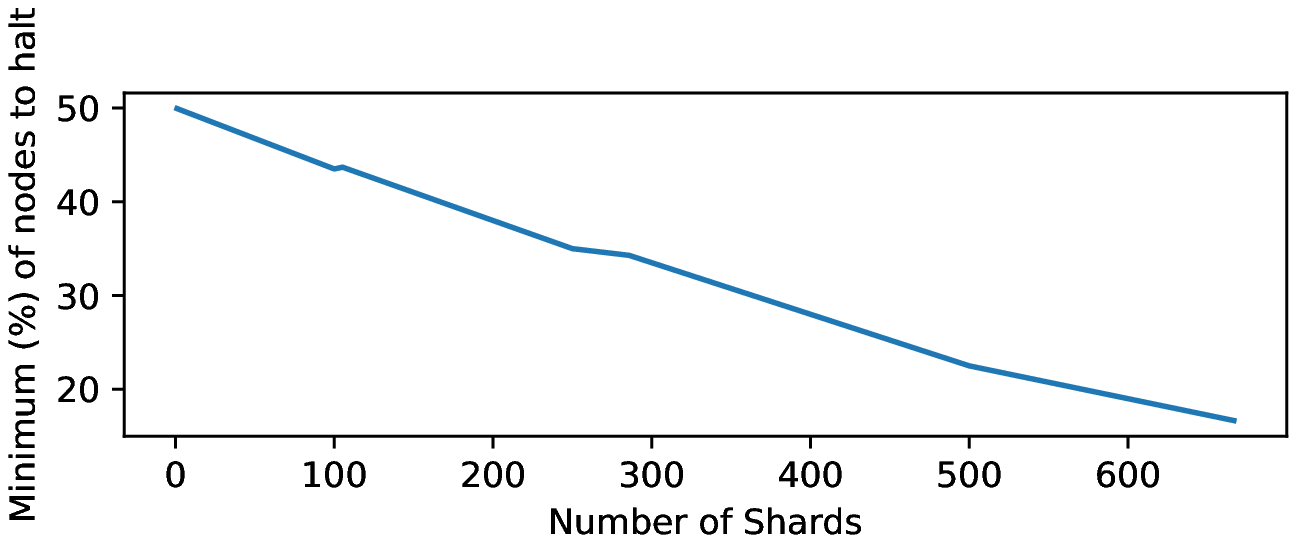}\\
{$n=1500$}&
	{$n=20000$}\\
\end{tabular}
	\caption{\fontfamily{lmr}\selectfont Minimum percentage of nodes required to halt a shard}
	\label{fig:img84-r1}
\end{figure}

In Figure \ref{fig:img8} we show the minimum percentage of nodes must be taken by the adversary to halt the entirety of the system with different number of nodes and different number of shards. 
\begin{figure}[ht!]
\centering
   \includegraphics[width=0.5\textwidth]{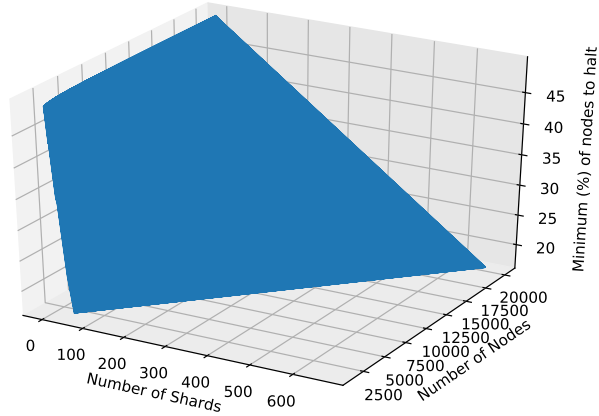}
	\caption{\fontfamily{lmr}\selectfont Minimum percentage of nodes required to halt the system as a whole}
	\label{fig:img8}
\end{figure}
As Figure \ref{fig:img8} suggests, with a fixed number of nodes, by reducing the number of shards, the minimum percentage is increased. Thus, using the halting recovery mechanism introduced in the model, the system will eventually recover from a halt when the number of shards reduced to the point that can defect the adversary. The honest will always win because the honest side has at least \emph{n/2+1} of nodes (the majority), thus meeting the same security level as the Nakamoto blockchain. As the rules discussed, \emph{T} in every shard is adjusted to keep a $10^{-6}$ failure chance assuming the adversary has \emph{n/2-1} of the nodes. This mechanism guarantees that the maximum chance to control a shard when the adversary has no more than $50\$$ of nodes is $10^{-6}$ with any possible number of shards existed.

In order to get a block accepted by the shard, this block must get $T$ approvals in $m$ nodes of this shard. In Figure \ref{fig:img6-ee}, we show $T/m$ in two systems of $1500$ and $20000$ nodes. For an adversary block to be accepted, the adversary must get $T$ nodes inside the shard.
In Figure \ref{fig:img6}, we show the percentage of nodes (T/M) the adversary must take in order to guarantee an adversary block to be approved.
\begin{figure}[ht!]
\centering
	\begin{tabular}{ll}
   \includegraphics[width=0.5\textwidth]{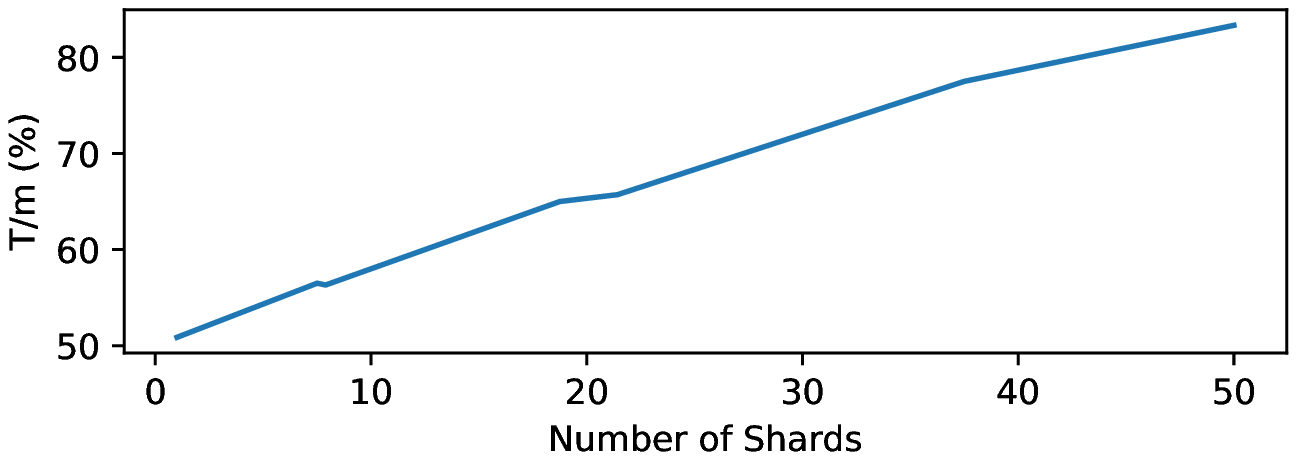}&\includegraphics[width=0.5\textwidth]{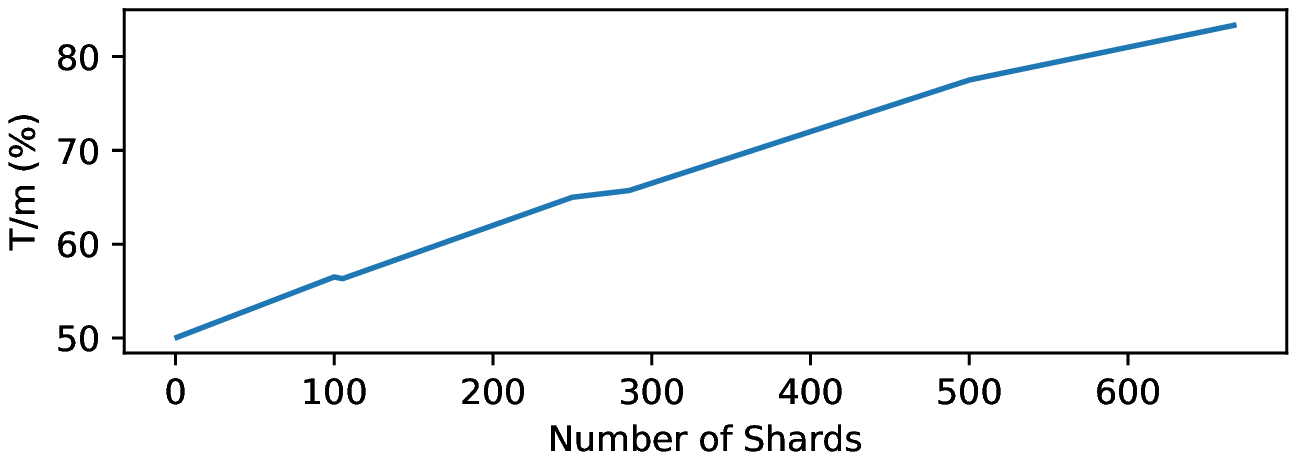}\\
{$n=1500$}&
	{$n=20000$}\\
\end{tabular}
	\caption{\fontfamily{lmr}\selectfont $T/M$}
	\label{fig:img6-ee}
\end{figure}
\begin{figure}[ht!]
\centering
   \includegraphics[width=0.5\textwidth]{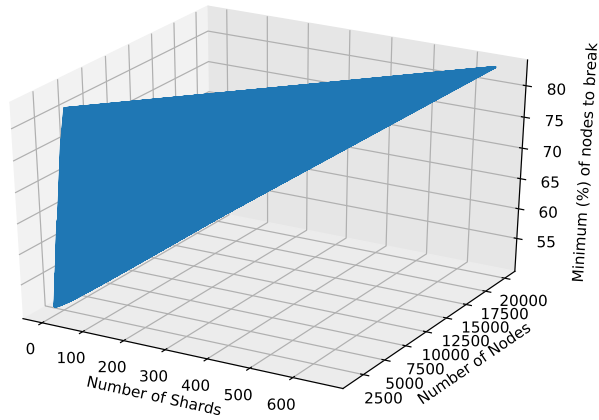}
	\caption{\fontfamily{lmr}\selectfont Minimum percentage of nodes required to guarantee controlling a shard}
	\label{fig:img6}
\end{figure}

Figure \ref{fig:img7-ee1} shows the transaction throughput per iteration with different shards in two systems of $1500$ nodes and $20000$ nodes (while fulfilling a $10^{-6}$ failure chance). Figure \ref{fig:img7} shows the transaction throughput per iteration with the different number of nodes and shards (while fulfilling a $10^{-6}$ failure chance). We let every shard block contain \emph{2000} transactions. 
\begin{figure}[ht!]
\centering
	\begin{tabular}{ll}
   \includegraphics[width=0.5\textwidth]{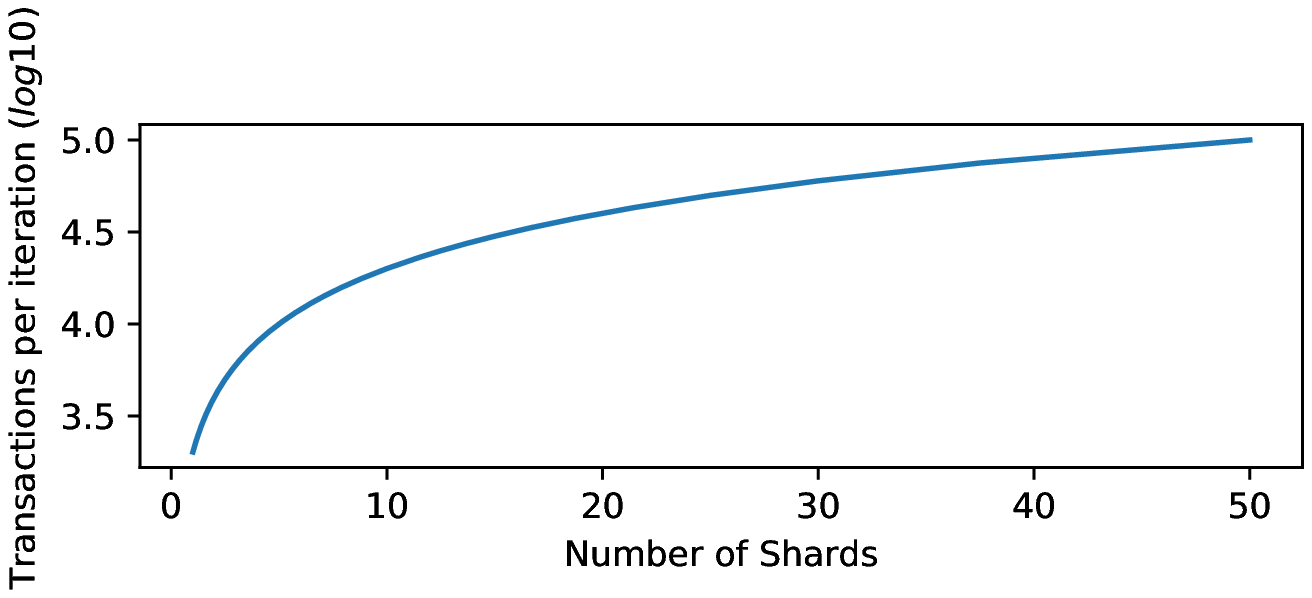}&\includegraphics[width=0.5\textwidth]{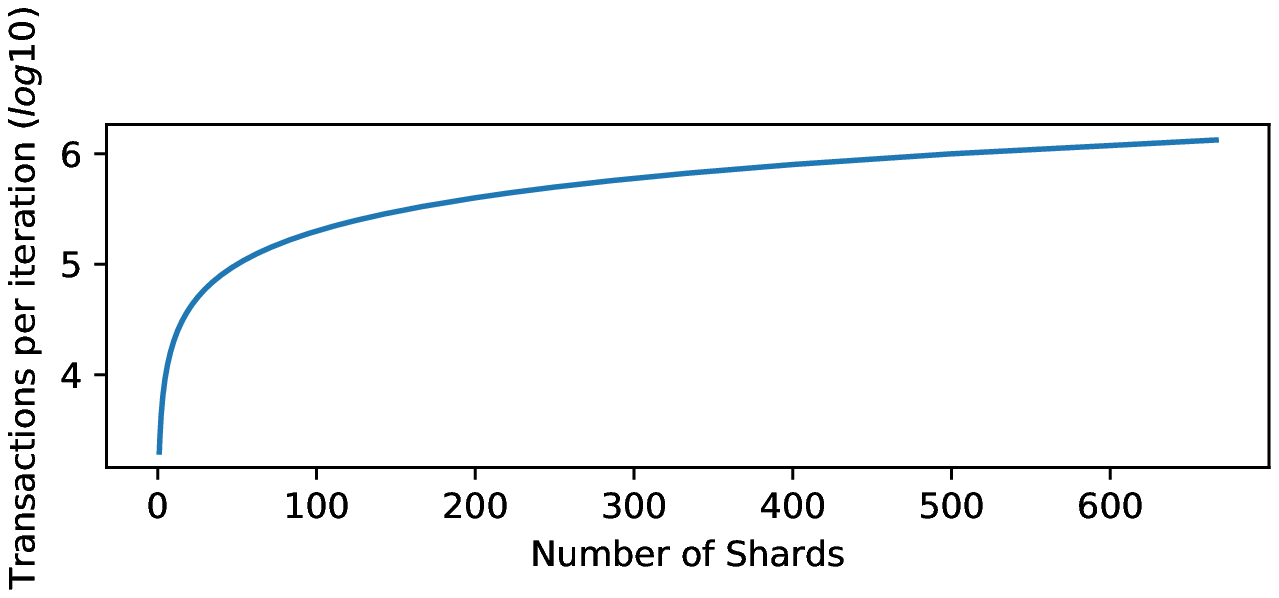}\\
{$n=1500$}&
	{$n=20000$}\\
\end{tabular}
	\caption{\fontfamily{lmr}\selectfont Transaction throughput per iteration with $2000$ transactions per block}
	\label{fig:img7-ee1}
\end{figure}
\begin{figure}[ht!]
\centering
   \includegraphics[width=0.5\textwidth]{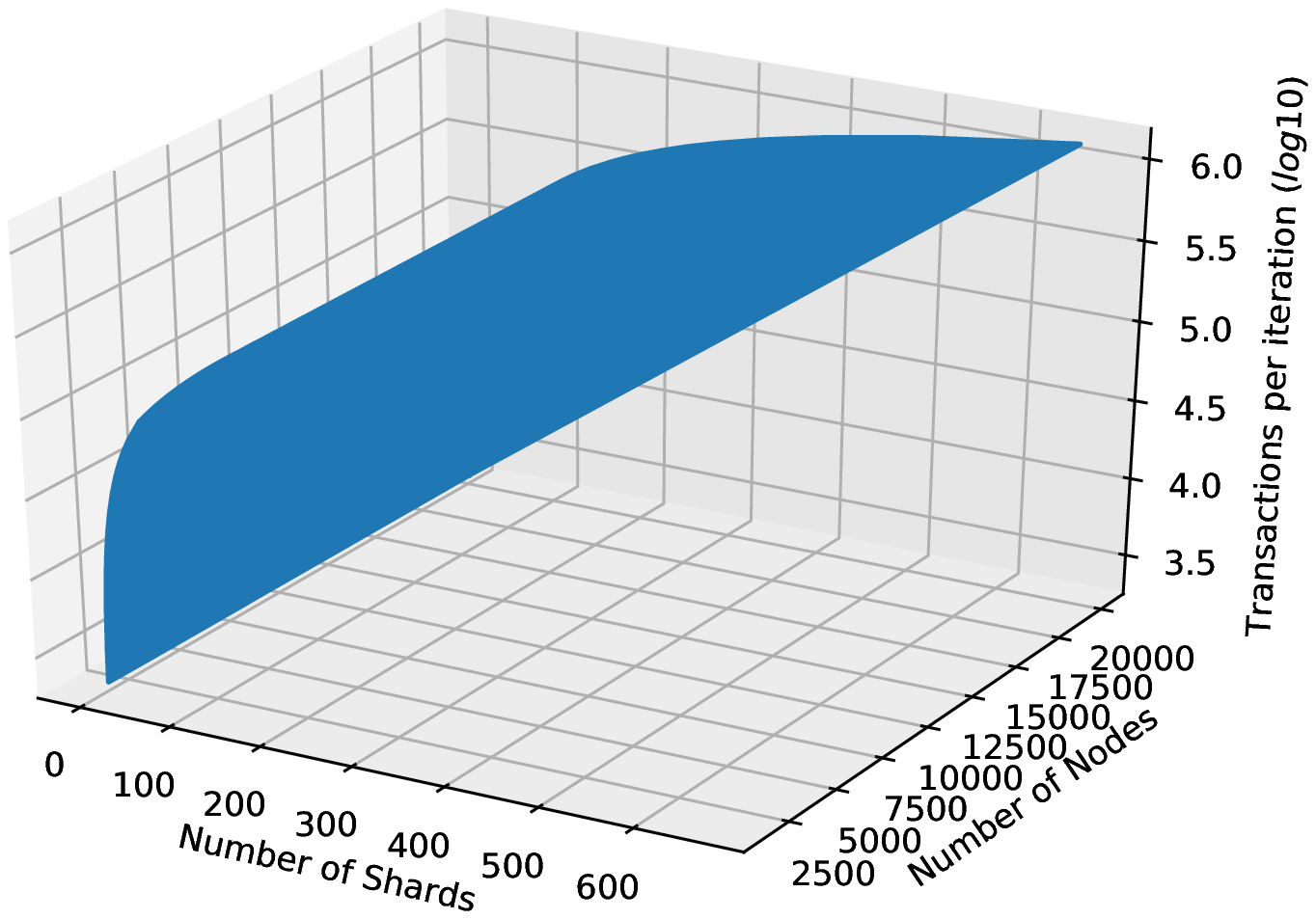}
	\caption{\fontfamily{lmr}\selectfont Transaction throughput per iteration with $2000$ transactions per block}
	\label{fig:img7}
\end{figure}

Let every block of a shard sized \emph{1Mbytes} (\emph{2000} transactions), every node signs one block per iteration. Let a block header sized \emph{150 bytes}, a transaction sized \emph{500 bytes} and a signature sized \emph{512 bytes} (a SHA256withECDSA signature and \emph{256 bytes} \emph{PoW} nonce). Let the committee's block $Size_{block_{committee}}\approx 33*n$ (contains all the \emph{33 bytes} public identity key of nodes). The quantified data requirement $DR$ in every iteration is:
\begin{equation}
    DR=Size_{block header} \times s + Size_{transaction} \times 2000 + Size_{signature} \times m + Size_{Block_{committee}} + Size_{Block_{Current\ shard}}
\end{equation}

Figure \ref{fig:img9-ee1} and Figure \ref{fig:img9} show the data requirement paired to the situation of Figure \ref{fig:img7-ee1} and Figure \ref{fig:img7}.
\begin{figure}[ht!]
\centering
	\begin{tabular}{ll}
   \includegraphics[width=0.5\textwidth]{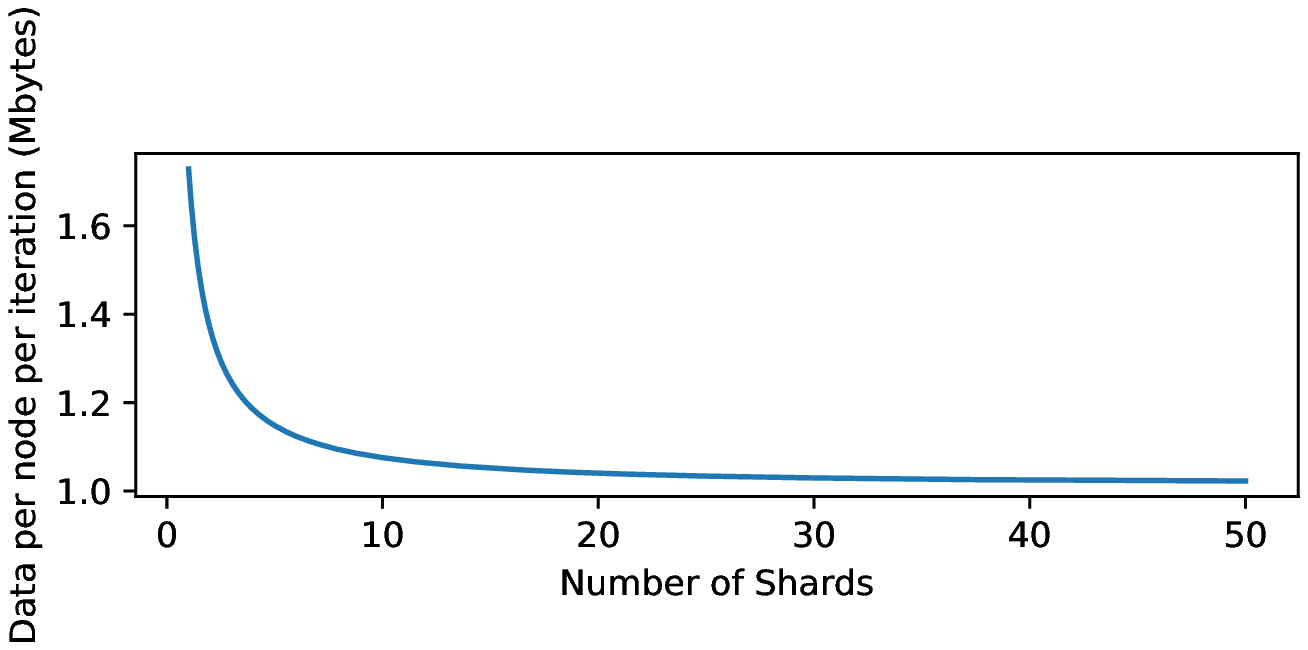}&\includegraphics[width=0.5\textwidth]{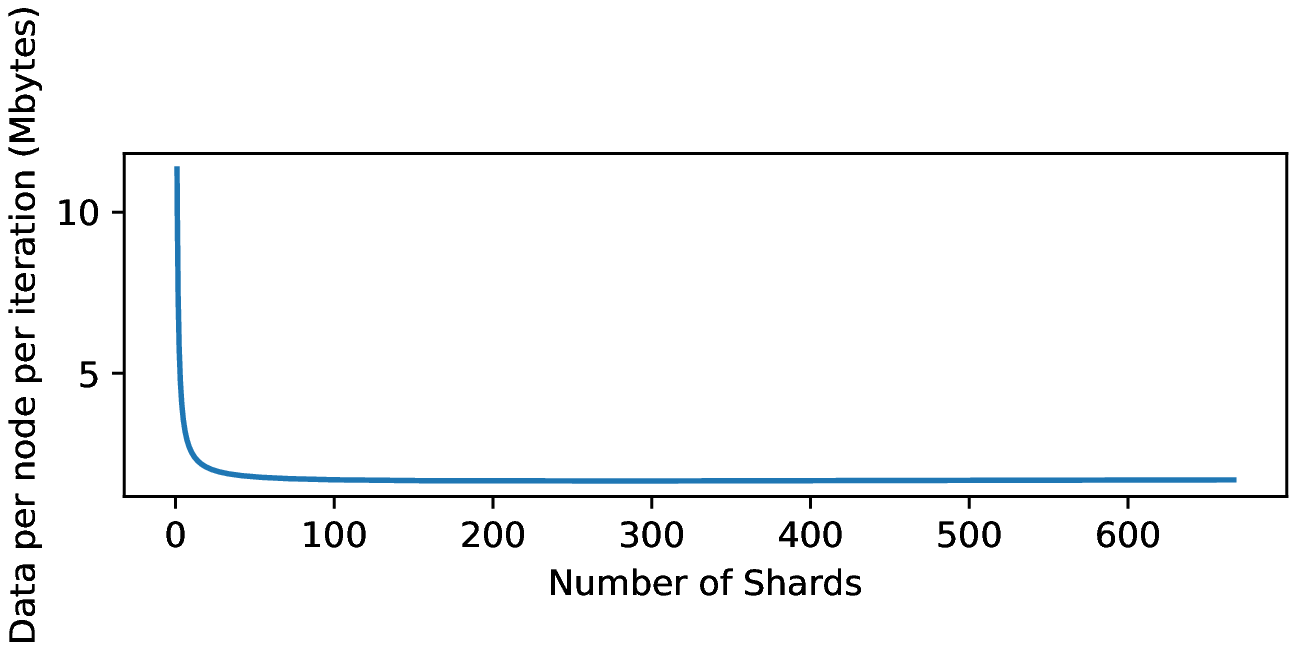}\\
{$n=1500$}&
	{$n=20000$}\\
\end{tabular}
	\caption{\fontfamily{lmr}\selectfont Data requirement}
	\label{fig:img9-ee1}
\end{figure}

\begin{figure}[ht!]
\centering
   \includegraphics[width=0.5\textwidth]{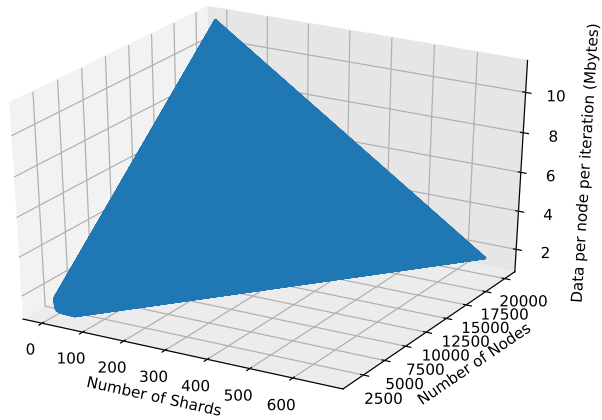}
	\caption{\fontfamily{lmr}\selectfont Data requirement per node per iteration}
	\label{fig:img9}
\end{figure}
\clearpage
\section{Conclusion}
In this paper, we showed an improved $n/2$ adversary node resistant Blockchain sharding approach; solved the problem of halting when the adversary has a lower than $n/2$ of nodes. In this way, making the Blockchain sharding approach reach both $n/2$ tampering resistance and $n/2$ halting resistant levels. It also makes the shard number become adjustable by the workload, and the workload per shard is balanced. With this improved approach, the performance of DAO systems can be primarily improved while the security of them is sustained. More devices can be added to the system, that extends the usage and potential applications of the blockchain. 

We admit that when applying large scale users to this approach, the adjustment of shards maybe frequent because of nodes can join in and leave the shards frequently. We believe such adjustment can be reduced by node regulation especially through the mean of financial tools \cite{xu2020anchoring}. In addition, when there are many shards, cross-shard communication can become an issue for the system and causing unnecessary verification or data exchange, the future work of this paper should focus on design and implementing cross-shard communication protocol. 

\bibliography{main}
}
\end{document}